\documentclass[a4paper,11pt]{article}
\usepackage{pos}

\newcommand{\be}{\begin{equation}}
\newcommand{\ee}{\end{equation}}

\newcommand{\half}{\frac{1}{2}}

\title{Reconstruction of bottomonium spectral functions in thermal QCD using Kernel Ridge Regression}
\ShortTitle{Bottomonium spectral functions using Kernel Ridge Regression}
\author*[a]{Sam Offler}
\author[a,b]{Gert Aarts}
\author[a]{Chris Allton}
\author[c]{Benjamin J\"ager}
\author[d]{Seyong Kim}
\author[e]{Maria Paola Lombardo}
\author[a]{Ben Page}
\author[f]{Sinead M. Ryan}
\author[c]{Jon-Ivar Skullerud}
\author[a]{Thomas Spriggs}

\affiliation[a]{Department of Physics, Swansea University, Swansea SA2 8PP, United Kingdom}

\affiliation[b]{European Centre for Theoretical Studies in Nuclear Physics and Related Areas (ECT*) \& Fondazione Bruno Kessler Strada delle Tabarelle 286, 38123 Villazzano (TN), Italy}

\affiliation[c]{CP3-Origins \& Danish IAS, Department of Mathematics and Computer Science, University of
Southern Denmark, 5230 Odense M, Denmark}

\affiliation[d]{Department of Physics, Sejong University, Seoul 143-747, Korea}

\affiliation[e]{INFN, Sezione di Firenze, 50019 Sesto Fiorentino (FI), Italy}

\affiliation[f]{School of Mathematics, Trinity College, Dublin 2, Ireland}

\affiliation[g]{Dept. of Theoretical Physics, National University of Ireland Maynooth, County Kildare, Ireland}

\emailAdd{s.p.offler.967106@swansea.ac.uk}

\abstract{
We discuss results for bottomonium at nonzero temperature obtained using NRQCD on {\sc Fastsum} {\em Generation 2L} ensembles, as part of the {\sc Fastsum} collaboration's programme to determine the spectrum of the bottomonium system as a function of temperature using a variety of approaches.
Here we give an update on results for spectral functions obtained using Kernel Ridge Regression. We pay in particular attention to the generation of training data and introduce the notion of using lattice QCD ensembles to learn how to improve the generation of training data. A practical implementation is given.
}

\FullConference{
 The 38th International Symposium on Lattice Field Theory, LATTICE 2021
  26th-30th July, 2021
  Zoom/Gather@Massachusetts Institute of Technology
}

\begin{document}
\maketitle

\section{Introduction}

Since the initial proposal that suppression of $J/\psi$ could be a signal for the formation of the quark-gluon plasma (QGP) \cite{Matsui:1986dk}, quarkonia, bound states of heavy quark-antiquark pairs, have been intensively studied. These states serve as important probes for the plasma created in heavy-ion collisions as they do not reach thermal equilibrium with the medium. Initial work focused on the charmonium system, but more recently the focus of both experimental \cite{CMS:2012gvv} and theoretical (see e.g.\ Refs.\ \cite{Rapp:2008tf,Larsen:2019zqv}) studies has shifted to bottomonium.

The purpose of this contribution is to continue the investigation of bottomonium at nonzero temperature by the {\sc Fastsum} collaboration \cite{Aarts:2010ek,Aarts:2011sm, Aarts:2014cda}, with the aim of determining masses and widths of ground and excited states, provided they exist, as well as the state-dependent melting or dissociation temperatures.
In Ref.\ \cite{Offler:2019eij} Kernel Ridge Regression (KRR) was introduced to reconstruct bottomonium spectral functions from Euclidean lattice correlators. KRR is a machine learning method that requires training data and here we revisit intricacies of generating this data. In particular we introduce the notion of using lattice QCD ensembles to learn how to improve the generation of mock data and demonstrate how this may be implemented. 

\section{Lattice details}

Given that the bottom quark mass scale $M$ lies above any other scale in the problem, including the temperature $T$ of the quark-gluon plasma, it can be integrated out to obtain standard non-relativistic QCD (NRQCD) \cite{Lepage:1992tx, Bodwin:1994jh}, also at nonzero temperature \cite{Burnier:2007qm}. We use the same formulation as in Refs.\ \cite{Aarts:2011sm, Aarts:2014cda}, i.e.\  terms up to $\mathcal{O}(v^4)$ are included, where $v$ is the heavy quark velocity in the bottomonium rest frame. 
In the relation between the bottomonium Euclidean correlator and the spectral function, terms involving $M/T$ are exponentially suppressed, leading to a relation with a simplified (``$T=0$'') kernel,
\begin{equation}
    G(\tau)  = \int^{\omega_{\rm max}}_{\omega_{\rm min}} \frac{d\omega}{2\pi}\, K(\tau, \omega) \rho(\omega),
    \qquad\qquad K(\tau,\omega) = e^{-\omega\tau}, 
    \label{eqn:spect_relation}
\end{equation}
where it is understood that an additive constant $\sim 2M$ has to be added to the lower limit of the integral, to make contact with  phenomenology in full QCD.

\begin{table}[b]
    \begin{center}
    \begin{tabular}{ccccccc}
    \hline
    $1/a_\tau$ [GeV] & $a_s$ [fm] & $\xi = a_s / a_\tau$  & $N_s$  & $m_\pi$ [MeV] & $m_\pi L$ & $T_{\rm pc}$ [MeV] \\
    \hline
    $5.997(34)$ & $0.01136(6)$ & $3.453(6)$ & $32$  & $236(2)$ & $4.36$ & $164(2)$ \\
    \hline
    \end{tabular}
    \caption{
        Parameters relevant for the ensembles: $a_s$ ($a_\tau$) is the spatial (temporal) lattice spacing; $\xi$ is the renormalised anisotropy; $N_s$ is the number of points in the spatial direction; $m_\pi$ is the mass of the pion; $T_{\rm pc}$ is the pseudocritical temperature, obtained from the inflection point of the renormalised chiral condensate~\cite{Aarts:2020vyb}.    
    }
    \label{tab:1}
\end{center}
\end{table}

For the finite-temperature study, we use the anisotropic {\sc Fastsum} {\em Generation 2L} ensembles, with $N_f=2+1$ flavours of clover-improved Wilson fermions, see Ref.\ \cite{Aarts:2020vyb}. The light quarks are heavier than in nature, but the strange quark mass is at its physical value. Relevant parameters are given in Table \ref{tab:1}.  
In the fixed-scale approach, the temperature is varied by changing $N_\tau$, using the relation $T=1/(N_\tau a_\tau)$.  
The $N_\tau$ values and corresponding temperatures are shown in Table \ref{tab:temperatures}.

\begin{table}[t]
 \centering
    \begin{tabular}{c||c|c|c|c|c|c|c|c|c|c|c}
    $N_\tau$ & 128 & 64 & 56 & 48 & 40 & 36 & 32 & 28 & 24 & 20 & 16 \\
    \hline
    T [MeV] & 47 & 94 & 107 & 125 & 150 & 167 & 187 & 214 & 250 & 300 & 375 
    \end{tabular}
    \caption{Temporal lattice sizes and temperatures for the {\sc fastsum} Generation 2L ensembles \cite{Aarts:2020vyb}. }
    \label{tab:temperatures}
\end{table}

\section{Data generation}

 As is the case with many machine learning methods, Kernel Ridge Regression (KRR) requires training in order to make predictions. In the current context, this means it is necessary to generate a set of mock spectral functions and calculate the corresponding correlators according to Eq.~(\ref{eqn:spect_relation}), yielding the training set $\{\rho_i(\omega), G_i(\tau)\}$ ($i=1, \ldots, N_{\rm train}$).
 The importance of generating appropriate training data cannot be understated; it is only at this stage that we can incorporate any constraints/information we know of. 
 
 In this study we choose to construct mock spectral functions from the combination of $N_p=5$ Gaussian peaks, such that a single spectral function is written as
 \begin{equation}
     \rho(\omega) = \sum^{N_p}_{p=1} Z_p \exp\left(-\frac{(\omega- m_p)^2}{\Gamma_p}\right).
     \label{eqn:mock_spectrl}
 \end{equation}
Here $Z_p$ is the amplitude, $m_p$ is the position and $\Gamma_p$ is related to the width of the Gaussian peak. Other mock functions can be constructed; in Ref.\ \cite{Offler:2019eij} the logarithm of the spectral function was expanded in an orthonormal set of incomplete basis functions.  Continuing with the Gaussian peaks here, we generate a collection of peaks by sampling values for $\{Z_p, m_p, \Gamma_p\}$ from a set of distributions. 
Masses are chosen from the following exponential distribution,
\begin{equation}
    m_{\mathrm{GeV}} = 9 + \frac{1}{\zeta}e^{-\omega/\zeta},
\end{equation}
where $\zeta$ determines the decay rate of the distribution. These values are then converted to lattice units using
\begin{equation}
    m_{\mathrm{lat}} = a_\tau \left(m_{\mathrm{GeV}} - \Delta M\right),
    \end{equation}
where $\Delta M = 7.465$ GeV is the additive constant mentioned below Eq.~(\ref{eqn:spect_relation}). As in Refs.~\cite{Aarts:2011sm, Aarts:2014cda}, it is determined by equating the calculated zero-temperature ground state mass in the $\Upsilon$ channel to its experimental value.  Widths are determined by writing $\Gamma = 10^{-x}$ and selecting $x$ from a uniform distribution, $0<x<6$. The amplitudes were selected for a uniform distribution, $0< Z< 10$. 
We put $\rho(\omega) = 0 $ for $\omega < \omega_{\rm min}$ and $\omega > \omega_{\rm max}$. 
Finally, the mock spectral functions (\ref{eqn:mock_spectrl}) are normalised, using the relation at $\tau = 0$, 
\begin{equation}
    G(0) = \int^{\omega_{\rm max}}_{\omega_{\rm min}} \frac{d\omega}{2\pi}\, K(0,\omega)\rho(\omega) = 
    \int^{\omega_{\rm max}}_{\omega_{\rm min}} \frac{d\omega}{2\pi}\, \rho(\omega),
\end{equation}
where $G(0)$ is the source in the NRQCD formulation.
In total we generated a set of  20\,000 mock spectral functions, each consisting of $N_p=5$ peaks and characterised by $3N_p=15$ parameters. Only $N_{\rm train}=15\,000$ functions from this dataset were used for training the KRR model.

\begin{figure}[t]
    \centering
    \includegraphics[width=0.49\textwidth]{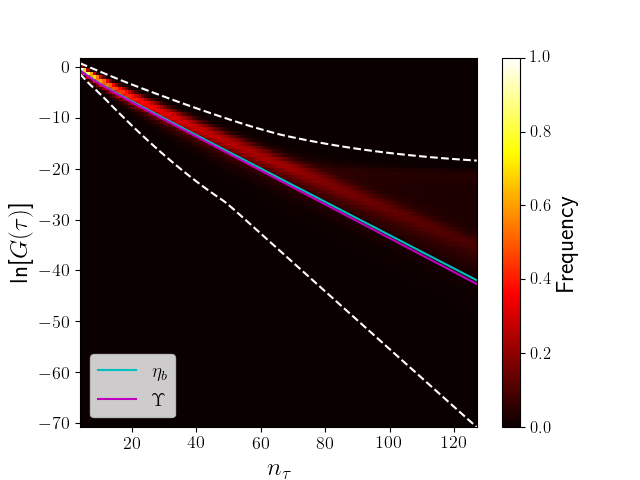}
    \includegraphics[width=0.49\textwidth]{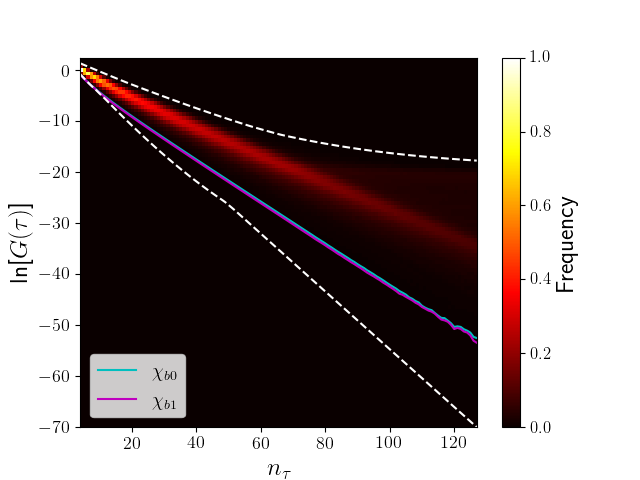}
    \caption{Distribution of the natural logarithm of correlators in the training set at each timeslice for $N_\tau = 128$ for S-wave (left) and P-wave (right) states.  Dashed lines represent the upper and lower limits of the correlators in the training set for each value of $n_\tau$. Solid lines represent the Euclidean correlators obtained in simulations. Note that the two channels in each figure -- $\eta_b$,  $\Upsilon$ (left) and $\chi_{b_0}$, $\chi_{b_1}$ (right) -- are hardly distinguishable. 
    }
    \label{fig:old-heatmaps}
\end{figure}

To assess whether the training set is potentially relevant for the correlators obtained from the actual lattice QCD simulations, we carry out the following comparison. At each timeslice we determine the distribution of (the logarithm of) the correlators in the training set, as obtained from the mock spectral functions. Given that the mock set contains 20\,000 spectral functions, the density quickly falls below 10\%. This distribution is shown in Fig.\ \ref{fig:old-heatmaps} via a heat map. The dashed lines represent the upper and lower limits of the correlators in the training set for each value of $n_\tau$.   
In addition we show the actual correlators obtained in the simulations. Both the left and the right panel in  Fig.\ \ref{fig:old-heatmaps} show two correlators -- $\eta_b$,  $\Upsilon$ (left) and $\chi_{b_0}$, $\chi_{b_1}$ (right) --, but note that these are hardly distinguishable. 

The key observation is that the training data compares poorly to the actual correlators, i.e.\ there is little overlap between the most common values seen in the training set and the real data. This effect is worse in the P-wave channels. Since machine learning algorithms assume that training and validation data sets are drawn from the same ensembles as (or at least are representative of) the real data sets, this is not a desired feature. 
We remark that the spread of the training data itself is not a problem; it is beneficial for KRR to have access to outliers in principle.

In order to generate a better set of training data, we use the actual correlators to learn which members of the training set are the most representative. This is implemented as follows: from the original training set, a subset of correlators are chosen by determining which correlator $G_i(\tau)$ lies closest to the actual correlator $G_{\rm NRQCD}(\tau)$, at each value of the temperature and for S-waves ($\eta_b$,  $\Upsilon$) and P-waves ($\chi_{b_0}$, $\chi_{b_1}$) separately. Closeness is here defined via
\be
    i = \mathrm{argmin}\sum^{N_{\tau}-1}_{n_\tau = 4}\left| \mathrm{ln}\left[\frac{G_i(n_{\tau})}{ G_{\rm{NRQCD}}(n_{\tau})}\right]\right|.
\ee
Note that the sum starts at $n_\tau=4$; the first few timeslices are not included in the summation. 
Given that there are ten different temperatures, see Table \ref{tab:temperatures}, and two S-wave or P-waves channels, this procedure yields a subset of up to 20 representative correlators (note that repeats are possible).  

This subset is used to construct a larger, more representative, training set, as follows. For each correlator in the subset, the corresponding spectral function and, more importantly, the set of $3N_p$ parameters \{$Z_p, m_p, \Gamma_p$\} are known. From those values the mean and variance is computed and new parameter values are sampled from Gaussian distributions with these means and variances. Note that for the amplitude $Z$ and the parameter $\Gamma$ this is done indirectly, using 
$\Gamma = 10^{-x}$ and an equivalent relation for $Z$, as above.
From this point onwards the procedure is the same as earlier, but with different distributions for the parameters characterising the mock spectral functions.  

\begin{figure}[t]
    \centering
    \includegraphics[width=0.49\textwidth]{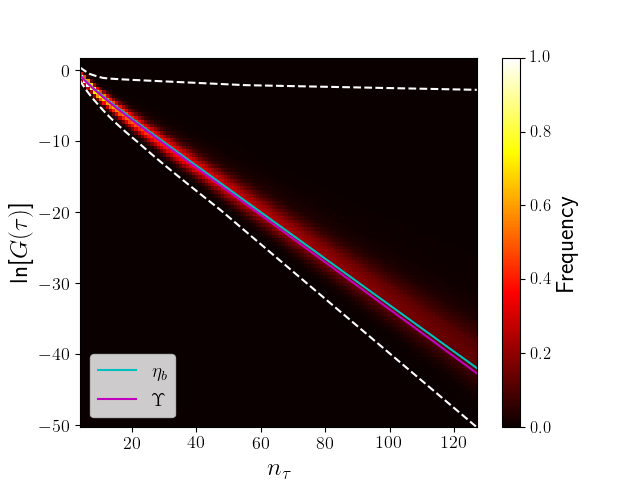}
    \includegraphics[width=0.49\textwidth]{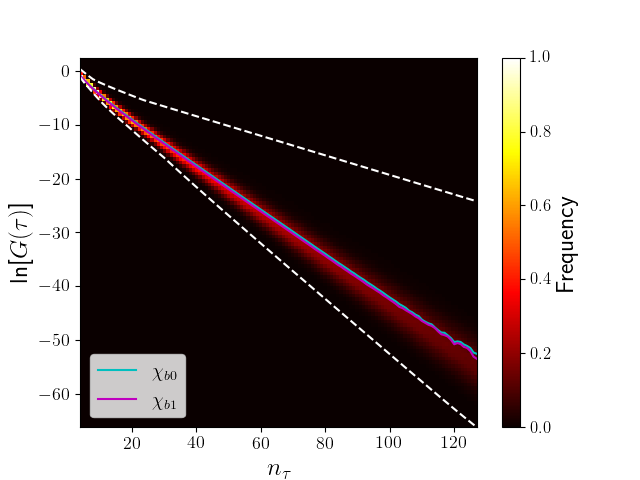}
    \caption{As above, for the improved training set.  }
    \label{fig:new-heatmaps}
\end{figure}

\begin{figure}
    \centering
    \includegraphics[width=0.33\textwidth]{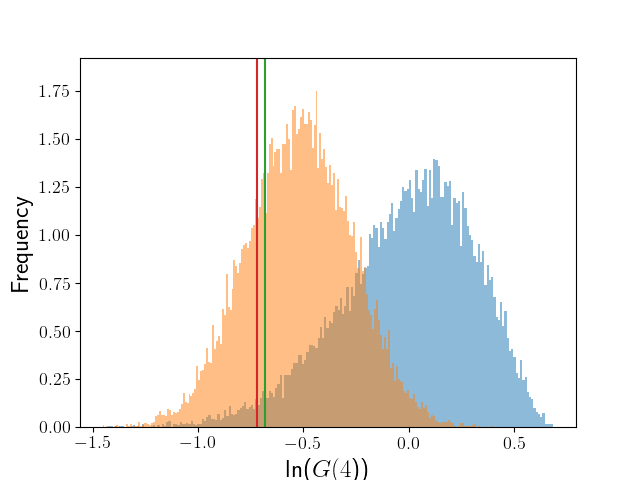}
    \!\!\!\!\!
    \includegraphics[width=0.33\textwidth]{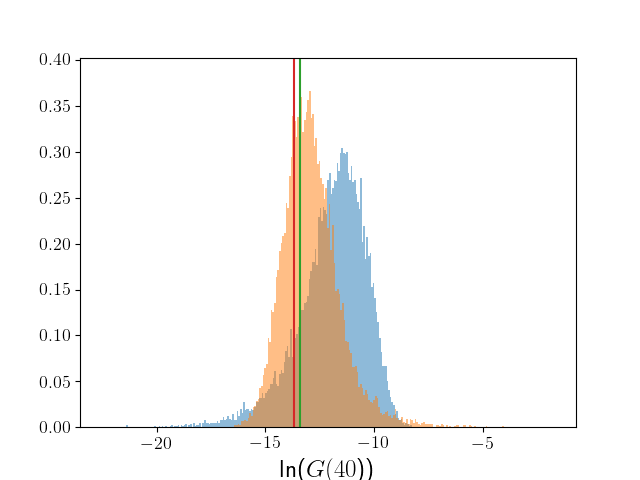}
    \!\!\!\!\!
    \includegraphics[width=0.33\textwidth]{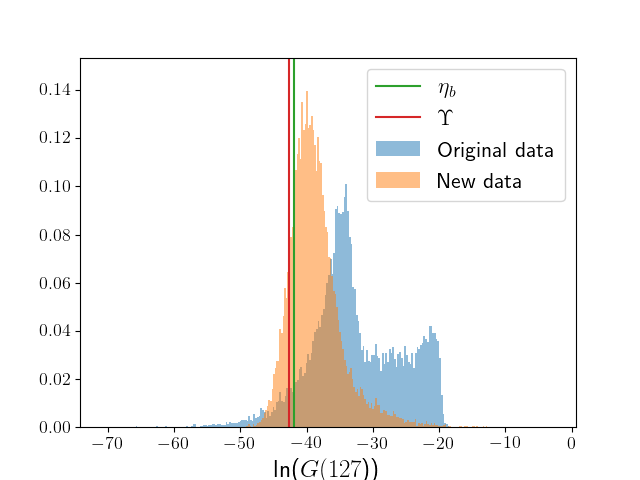}
    \caption{Distribution of the original and the improved training set for the correlators at $n_\tau$ = 4 (left), 40 (centre), and 127 (right) for $N_\tau = 128$. Solid vertical lines represent the value of the actual Euclidean correlators at the specified timeslices. }
    \label{fig:Timeslice_hist}
\end{figure}

Fig.\ \ref{fig:new-heatmaps} displays the same comparison as in Fig.\ \ref{fig:old-heatmaps}, but now for the improved training set. As can be seen, the most common correlators in the training set now coincide with the actual correlators, leading potentially to a more appropriate training set. 
This is further demonstrated in Fig.\ \ref{fig:Timeslice_hist}, where the distribution of correlators obtained from the original and the improved training sets are shown at selected timeslices for the $N_{\tau} = 128$ lattice. The actual correlator values at the selected timeslices are shown with the vertical lines. We indeed observe a better overlap between the actual correlators and the data in the training set. The latter still has a nonnegligible width, allowing for flexibility in the application to spectral reconstruction.

\section{Kernel Ridge Regression}

Kernel Ridge Regression (see e.g.\ Ref.\ \cite{bishop}) combines two techniques: the kernel method and ridge regression. Consider first linear regression in which $y = w^T\phi(x)$, where $y$ are the target data, $\phi(x)$ is a vector of functions of the input data $x$, and $w$ is a vector of parameters to be determined. The cost function to be minimized is $E=\half[y - w^T\phi(x)]^2$. 
In our application, the input data are the training set of Euclidean correlators $G_i(\tau)$ ($i=1,\ldots, N_{\rm train}$). However, rather than using these directly, they are used to generate an $N_{\rm train}\times N_{\rm train}$ matrix $\mathbf{C}$ (kernel function), with matrix elements  
\be
C_{ij} = \exp \left( -\gamma \sum^{N_{\tau}-1}_{n_\tau = 4} \left[\widetilde G_i(n_\tau) - \widetilde G_j(n_\tau)\right]^2\right),
\qquad\qquad
\widetilde G_i(n_\tau) = \frac{G_i(n_\tau)}{\overline{G}(n_\tau)}.
\ee
Note that the form of the kernel function is not unique.
Here $\gamma$ is a hyper-parameter, which sets a correlation length in the space of correlators. The correlator data appearing is normalised with $\overline{G}(n_\tau) = \sum_i G_i(n_\tau)/N_{\rm train}$, the mean of the training data at each timeslice. This is done to account for the difference in absolute size of the correlators as $\tau$ changes.

The target data in our application are the spectral functions. To control the size of the target data, we use here expansion (\ref{eqn:mock_spectrl}) in terms of Gaussian peaks, leading to $N_a \equiv 3N_p=15$ parameters for spectral function. The target data is then represented by a $N_{\rm train}\times N_a$ matrix $\mathbf{Y}$, in which each row contains the parameters of a single spectral function. It is now assumed that the input and target data are related as
\be
\mathbf{Y} = \mathbf{C}\mathbf{\alpha},
\ee
in which the $N_{\rm train}\times N_a$ matrix $\mathbf{\alpha}$ is the analogue of the vector $w$ in linear regression. The aim of the training stage is to determine the matrix $\mathbf{\alpha}$.

To prevent overfitting, it is common to add an additional term in the cost function, proportional to the square of the parameters (ridge regression). Hence the cost function to be minimised reads
\be
  E(\mathbf{Y, C, \alpha} ) = \half(\mathbf{Y} - \mathbf{C}\mathbf{\alpha})^2 + \half\lambda\mathbf{\alpha}^T\mathbf{C}\mathbf{\alpha},
\ee
where $\lambda$ is the second hyper-parameter, used to regularise the influence of the additional term. Minimising this cost function with respect to $\mathbf{\alpha}$ then determines the optimal parameter matrix, for given $\sigma$ and  $\lambda$, as
 \be
\mathbf{\alpha}_{\rm opt} = \left( \mathbf{C} + \lambda \mathbf{I} \right)^{-1} \mathbf{Y}.
\ee
The hyper-parameters are determined via a cross-validation procedure, see below.

After this training stage, it is possible to make predictions for a spectral function ($\mathbf{Y'}$) given an actual Euclidean correlator using
\be
\mathbf{Y'} = \mathbf{C'}\mathbf{\alpha}_{\rm opt},
\ee  
where $\mathbf{C'}$ is determined from the squared rescaled difference between the actual correlator and the training correlators, i.e.\ it is a matrix of size $1\times N_{\rm train}$.

\section{Optimization}

\begin{figure}[t]
    \centering
    \includegraphics[width=0.51\textwidth]{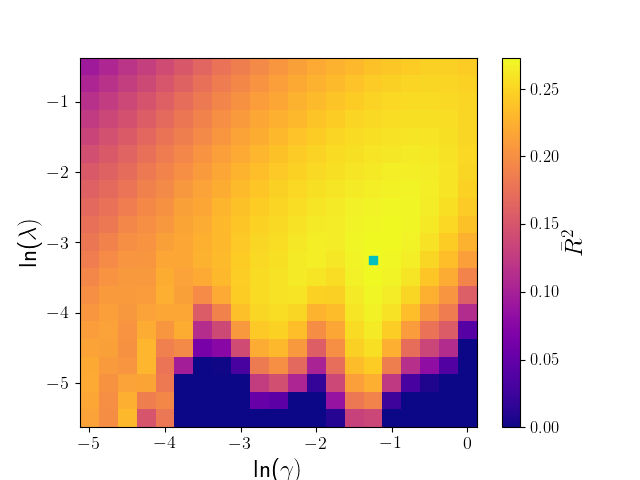}
    \!\!\!\!\!\! \!\!\!\!\!\!
    \includegraphics[width=0.51\textwidth]{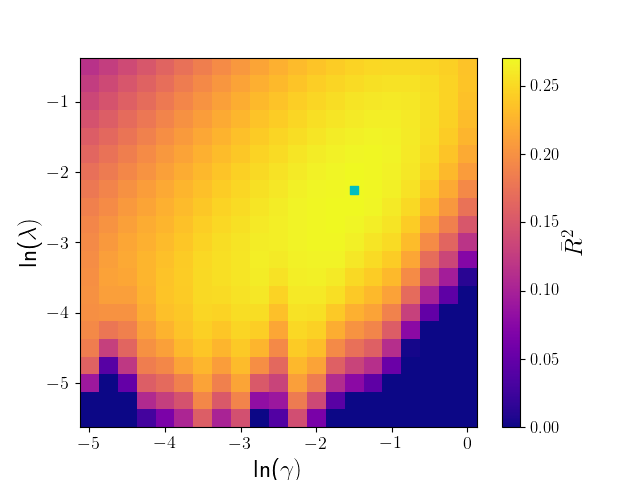}
    \caption{Heatmap of $\overline R^2$ used for cross-validation in the plane of hyper-parameters ($\lambda, \gamma$), 
    for $N_{\tau} = 20$ (left) and $N_{\tau} = 128$ (right). The maxima, indicated with the little squares, provide the optimal choice of hyper-parameters. In cases where $\overline R^2$ <0, it has manually been  set equal to 0 for clarity.}
    \label{fig:r_squared heatmaps}
\end{figure}

To determine the optimal choice of hyper-parameters, we consider the so-called $R^2$ score, the coefficient of determination, defined in general as \cite{R2}
\be
R^2(y) = 1 - \frac{\sum_i(y^{\rm true}_i - y^{\rm pred}_i)^2}{\sum_i (y^{\rm true}_i - \overline{y}^{\rm true})^2}.
\ee 
In  our application, $y$ is one of the $N_a=15$ parameters parametrising a spectral function and we consider the mean $R^2$ score, averaged over the parameters, 
\be
\overline{R}^2 = \frac{1}{N_a}\sum_{a=1}^{N_a} R^2(y_a).
\ee
Rather than using a training and a validation set, we use cross-validation \cite{bishop} in which a single set is used for both training and testing. The optimisation procedure seeks to find the value of $\overline R^2$ closest to 1.
Results for $\overline R^2$ are shown in Fig.\ \ref{fig:r_squared heatmaps}, for all the hyper-parameters pairs tested, and for two ensembles, $N_{\tau} = 20$ (left) and $N_{\tau} = 128$ (right). 
Though the plots are similar, there does appear to be some thermal dependence. In both cases there is a broad region of hyper-parameter pairs that perform to a similar standard. 
The most optimal choice of hyper-parameters is shown by the little squares, with $\left(\ln(\gamma), \ln(\lambda) \right)$ = (-1.25, -3.25) (left) and $\left(\ln(\gamma), \ln(\lambda)\right)$ = (-1.5, -2.25) (right).
It is noted that the optimal values of $\overline R^2$ are rather low, i.e.\ not close to 1.
Looking at the $R^2$ scores for each of the $N_a$ parameters individually, we note that the $R^2$ score for the ground state mass is $R^2_{m_0} = 0.869$, whereas the $R^2$ scores for the groundstate width and amplitude are significantly lower, $0.148$ and $0.336$ respectively. We may therefore further improve the procedure by treating the various parameters on unequal footing.

\begin{figure}[t]
    \centering
    \includegraphics[width=0.7\textwidth]{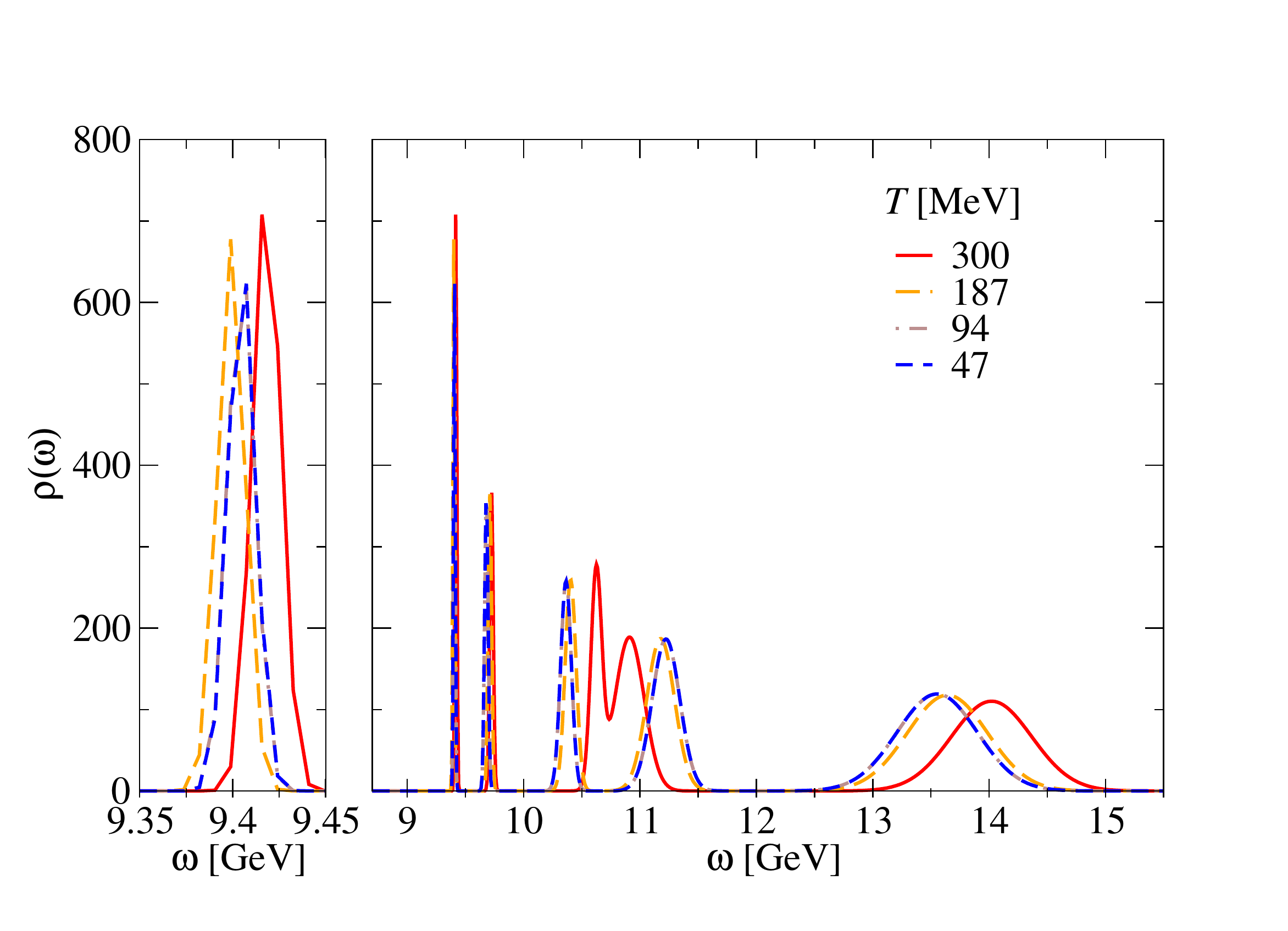}
    \caption{Spectral functions in the $\Upsilon$ channel at 4 different temperatures, obtained using Kernel Ridge Regression (KRR). The left pane zooms in on the ground state peak.
        }
    \label{fig:Spectral_preds}
\end{figure}

As a preliminary result, Fig.\ \ref{fig:Spectral_preds} finally presents the predicted spectral functions in the $\Upsilon$ channel at four different temperatures. It is noted that the masses of both the ground  and first excited state are smaller than the expected values. This may be due to the choice of expressing the spectral functions in terms of $N_p=5$ Gaussian peaks and requires further investigation. 
Returning to the representation described in \cite{Offler:2019eij} may improve this. For the application of alternative methods to the same data, see Refs.\ \cite{tom_lat21, ben_lat21}, also presented at this conference. 

\section{Summary}

The main message of this contribution is that it is possible to adapt the training set by comparing its predictions to the actual correlator data from lattice QCD simulations, at the level of the correlators. A large discrepancy would suggest that the training sets are not representative of the real data sets, an undesired feature. It is noted that improving the overlap is necessary, but not sufficient.
As a general remark, we note that this method can be iterated in principle and is applicable to other parametrisations of training spectral functions as well. The parametrisation used here -- spectral functions are written as sums of Gaussian peaks -- may well be an important limiting factor.

\acknowledgments 

We thank  Jonas Glesaaen for collaboration at the early stages of this project.
This work is supported by STFC grant ST/T000813/1.
SK is supported by the National Research Foundation of Korea under grant NRF-2021R1A2C1092701 funded by the Korean government (MEST).
BP has been supported by a Swansea University Research Excellence Scholarship (SURES).
This work used the DiRAC Extreme Scaling service at the University of Edinburgh, operated by the Edinburgh Parallel Computing Centre on behalf of the STFC DiRAC HPC Facility (www.dirac.ac.uk). This equipment was funded by BEIS capital funding via STFC capital grant ST/R00238X/1 and STFC DiRAC Operations grant ST/R001006/1. DiRAC is part of the National e-Infrastructure.
This work was performed using PRACE resources at Cineca via grants 2015133079 and 2018194714.
We acknowledge the support of the Supercomputing Wales project, which is part-funded by the European Regional Development Fund (ERDF) via Welsh Government,
and the University of Southern Denmark for use of computing facilities.
We are grateful to the Hadron Spectrum Collaboration for the use of their zero temperature ensemble.

\end{document}